\documentclass[10pt,journal,compsoc]{IEEEtran}
\usepackage{etoolbox}
\usepackage{lipsum}
\usepackage{booktabs}
\usepackage{amsmath, amssymb}
\usepackage{xcolor}
\usepackage{hyperref}
\usepackage[noadjust]{cite}
\usepackage{filecontents}
\usepackage{graphicx}
\usepackage{graphicx}
\usepackage{tabularx,booktabs} 
\usepackage{subfigure}%
{}
{}
{}
\usepackage{longtable}
\usepackage{longtable,pdflscape,booktabs}
\usepackage{subfigure}
\usepackage{algorithm}
\usepackage{algpseudocode}
\usepackage{lineno, blindtext, color}
\usepackage[skip=1pt]{caption}
\makeatletter
\algrenewcommand\ALG@beginalgorithmic{\footnotesize}
\makeatother

\ifCLASSOPTIONcompsoc
\else
  \usepackage{cite}
\fi

\ifCLASSINFOpdf
\else
\fi

\begin{document}

\title{A Review of Performance, Energy and Privacy of Intrusion Detection Systems for IoT}
\author{Junaid Arshad, Muhammad Ajmal Azad, Khaled Salah, Wei Jie, Razi Iqbal, Mamoun Alazab
\IEEEcompsocitemizethanks{\IEEEcompsocthanksitem Junaid Arshad is with University of West London, UK, Muhammad Ajmal Azad is with The Warwick University, UK,  Khaled Salah is with Khalifa University, UAE, Wei Jie is with University of West London, UK, Razi Iqbal is with College of Computer Information Technology, American University in the Emirates, UAE, Mamoun Alazab is with Charles Darwin University, Australia.\protect\\

}

}

{}
\IEEEtitleabstractindextext{%
\begin{abstract}
Internet of Things (IoT) is a disruptive technology with applications across diverse domains such as transportation and logistics systems, smart grids, smart homes, connected vehicles, and smart cities. 
Alongside the growth of these infrastructures, the volume and variety of attacks on these infrastructures has increased highlighting the significance of distinct protection mechanisms.
Intrusion detection is one of the distinguished protection mechanisms with notable recent efforts made to establish effective intrusion detection for IoT and IoV. However, unique characteristics of such infrastructures including battery power, bandwidth and processors overheads, and the network dynamics can influence the operation of an intrusion detection system. This paper presents a comprehensive study of existing intrusion detection systems for IoT systems including emerging systems such as Internet of Vehicles (IoV). The paper analyzes existing systems in three aspects: computational overhead, energy consumption and privacy implications. Based on a rigorous analysis of the existing intrusion detection approaches, the paper also identifies open challenges for an effective and collaborative design of intrusion detection system for resource-constrained IoT system in general and its applications such as IoV. These efforts are envisaged to highlight state of the art with respect to intrusion detection for IoT and open challenges requiring specific efforts to achieve efficient intrusion detection within these systems. 
\end{abstract}

}

\maketitle

\IEEEdisplaynontitleabstractindextext

\IEEEpeerreviewmaketitle

\section{Introduction}
\label{sec:introduction}
The use of sensor devices has witnessed an extraordinary increase over the last few years leading to their proliferation across diverse domains such as wearables, intelligent appliances, and vehicles. The ability of these devices to be connected to a network has introduced exciting new paradigms such as the Internet of Things (IoT). IoT has received significant attention as a disruptive technology and is fundamental to the networks of the future. A recent study by Frost and Sullivan has predicted the number of IoT devices to increase to 45.41 billion devices in 2023 \cite{1}. This has caused direct impact on industrial applications such as automotive industry, commercial security cameras, as well as consumer applications such as wearables, smart TVs, and smart meters.


A typical IoT system consists of devices with resource constraints such as limited processing power, energy resources and communication range etc. These constraints mandate communication technologies that require limited energy overheads, provide efficient performance under diverse conditions and support larger address space. A number of advanced wireless communication technologies have emerged to facilitate these requirements. These include RFID \cite{RFID2017}, Wireless Sensor Networks (WSN) \cite{WSN2015}, Zig Bee \cite{Zigbee2015}, Bluetooth and IPv6 over Low power Wireless Personal Area Networks (6LoWPAN)\cite{2} and have shaped M2M  communication  as  well  as  dedicated  communication  technologies for emerging paradigms such as IoT and IoV.

However, the open network architecture of IoT 
has also attracted intruders to use this network of thousands of devices for spreading malicious content such as the recent IoT botnets \cite{Mirai2016,bricker,Reaper2018}. Furthermore, a recent study by Gartner \cite{gartner} has predicted IoT based attacks to form 25\% of all enterprise attacks by 2020 highlighting the need for distinct protection mechanisms. Due to the proliferation of such devices in almost every aspect of our life, the threats posed due to their insufficient security are unique with insecure devices exposing the end users to serious security and privacy threats. For instance, if an attacker is able to compromise an in-car WiFi; in-car devices and data will be at risk. Once inside the network, an attacker can spoof the car, connect to outside data sources, and steal the owner’s personal information including credit card data \cite{6}. 

In view of such emerging threats, the need to address security challenges for an IoT system is paramount. There have been a number of efforts to address different dimensions of security for IoT such as secure frameworks \cite{SF1, SF2}, privacy of information \cite{SF3}, and authentication \cite{SF4}. However, the challenge in the design of an effective secure system is two-fold: firstly, the devices which form these systems are typically resource constrained which limits their ability to host sophisticated security management system that can monitor the device in real time, secondly the ad-hoc nature of IoT and IoV systems allows devices to connect to other devices at runtime, typically for short time periods, thereby creating a collaborative network. \\
Intrusion detection systems (IDS) are normally placed at the edge of the network to strengthen security capabilities of a system in the event of a malicious attempts. Over the last few years, intrusion detection systems for the IoT system have received increased attention with a number of intrusion detection systems proposed such as \cite{LR1, LR16, LR28}. Within this context, the focus of our research is to investigate novel challenges to achieve efficient intrusion detection for IoT systems and explore dedicated solutions to address them. In view of this, we present an in-depth study of the state-of-the-art with respect to intrusion detection within IoT identifying limitations of current approaches and highlighting future directions.

Through our research, we have identified existing efforts to study the state-of-the-art for intrusion detection for IoT systems with \cite{S1} and \cite{S2} being the most notable efforts. Although these efforts present a structured analysis of existing literature within IoT intrusion detection domain, these share a significant limitation in being agnostic of the performance overhead and privacy implications. For instance, using the \textit{interaction ability} proposed by \cite{S1}, Jun and Chi \cite{R4} scores high (three) indicating its efficiency to protect a IoT system. However, a deeper analysis of the system highlights that it is remarkably CPU intensive which will affect its suitability for IoT system. In view of these limitations, we undertake rigorous analysis of the intrusion detection systems taking into account a number of performance related attributes to present an exhaustive evaluation of existing approaches for intrusion detection in IoT systems.
\\
Considering the limitations of the existing studies, this paper makes following contributions:
\begin{itemize}
   \item A comprehensive attack model for IoT systems which is envisaged to inform state of the art for intrusion detection within IoT. The attack model comprises of threats across different dimensions of an IoT system aiming to aid improved classification.
\item An extensive review of efforts with respect to intrusion detection for IoT systems. An in-depth review has been conducted taking into account critical attributes such as performance overhead and privacy implications.
\item Identification of open challenges to achieve effective intrusion detection for IoT systems. This is informed by extension review and analysis of existing intrusion detection research within IoT systems. 
\end{itemize}
Rest of the paper is organized as follows. Section \ref{sec:attackmodel} introduces a comprehensive attack model for IoT systems. Section \ref{sec:existingsurveys} includes a detailed discussion about the existing reviews of intrusion detection research within IoT highlighting unanswered questions. Extensive review of existing literature within intrusion detection for IoT is presented in Section \ref{sec:reviews} which is organized into different types of IDS and therefore provides a linkage with classification introduced in Section \ref{sec:existingsurveys}. Section \ref{sec:privacyimplications} presents privacy implications of intrusion detection systems followed by \ref{sec:analysis} which provides a thorough analysis of the existing literature with respect to a number of attributes including intrusion detection and performance efficiency. Through the findings of our research, section \ref{sec:challenges} discusses open challenges which require further attention followed by conclusions and future aims of our research in section \ref{sec:conclusions}.

\section{An Attack Model for IoT Systems}
\label{sec:attackmodel}
Although IoT is an emerging paradigm, a significant part of the software stack used by the IoT applications is adopted from existing software paradigms. This is also evident from the concept of integrating IoT specific stack (such as specific to Zigbee, 6LoWPAN and RPL) with the existing Internet infrastructure such as IPv5 and IPv6. This has significant implications with respect to the threat model for IoT infrastructures as it is not restricted to the threats specific to the new routing protocols such as 6LoWPAN and RPL but also includes threats to existing infrastructure such as IPv6, application specific attacks and attacks specific to the physical media such as the radio spectrum. In this section we present a taxonomy of attack types for a typical IoT system.

\begin{figure*}[t]
  \begin{center}
 \includegraphics[width=180mm,height=65mm]{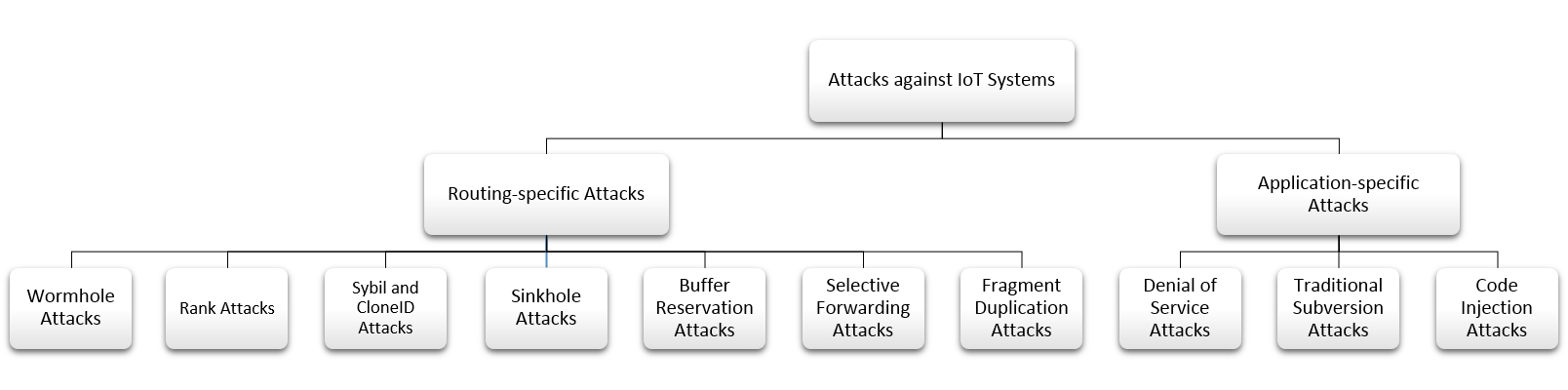}
 \end{center}
  \setlength{\belowcaptionskip}{-15pt}
  \caption{A Taxonomy of attacks for IoT systems}
  \label{fig:fig3}
  \end{figure*}

\subsection {Routing-specific threats}
Routing information in IoT system can be modified or spoofed in order to divert the traffic or create an attack on the IoT network. These attacks are most common attacks in the resource-constrained IoT and sensor networks. The most relevant routing attacks in IoT includes the following:

\textbf{\textit {Rank attack:}}
A defining characteristic of 6LoWPAN networks is the use of ranking to establish optimal routing path. Within this context, \textit{Node Rank} indicates the quality of the path from a node to the sink node. Every time a node updates its rank or preferred parent, it needs to inform other nodes by sending the updated information in the next DIO. RPL uses the Rank rule that \textit{a node in the parent should always have lower rank than its children to prevent the loop creation}. This way, the rank enables creating optimal topology, preventing loop creation and managing control overhead \cite{XR2}. As identified by \cite{XR2,XR3,XR4} the rank information can be maliciously tampered with by an attacker so that it chooses the node with worst Rank to be its parent. This will therefore result in disturbing the topology of the network causing delays in normal transmission.  

\textbf{\textit {Wormhole attack:}}
A wormhole can be considered as a tunnel between two nodes using wired or wireless links and can be used to achieve faster transmission rates or dedicated connection between such nodes. As such, wormhole has legitimate applications such as the connection between the local and global IDS modules within our architecture. Wormhole can be used by an attacker to create a dedicated tunnel with a node on the Internet as identified by \cite{XR5}. Wormhole attack is not novel to the IoT systems and has been historically identified as a potential threat for wireless sensor networks by \cite{XR6, XR7,XR8}.

\textbf{\textit {Sinkhole attack:}}
The objective of a sinkhole attack is to attract traffic through a designated node using illegitimate information making the node a lucrative routing sink (base station within wireless network terminology). As with wormhole attack, literature around sinkhole attack is well established with \cite{XR9} being an initial effort to identify and mitigate against such attack. Creating a sinkhole does not necessarily disrupt legitimate transmission within a 6LoWPAN however by diverting the traffic through a specific route creates opportunities to launch other attacks such as wormhole and selective forwarding attack described below.

\textbf{\textit {Selective forwarding attack:}}
With selective forwarding attack, a malicious node attempts to disrupt legitimate transmission and routing path. The malicious node in this case attempts to block certain packets and forward selected packets thereby affecting the routing. For instance, an attacker can forward all RPL control messages but block the rest \cite{XR5}. This attack can cause more damage when used in conjunction with sinkhole attack. Such dependencies among different attack types has motivated us to explore the impact of multi-stage attacks within IoT infrastructures. To the best of our knowledge the intrusion detection system presented in this paper is the pioneer effort to identify this issue and explore solution to mitigate against it.

\textbf{\textit {Fragment duplication attack:}}
The fragment duplication attack leverages a weakness within the 6LoWPAN layer with respect to how fragmented packets are received and assembled by an IoT node. A consequence of the integration of 6LoWPAN with IPv6 networks is that larger packets supported by IPv6 have to be fragmented into smaller packets so as to be effectively processed by the resource-constrained nodes within an IoT system. However, as identified by \cite{XR1}, a recipient node cannot verify if two fragments of a packet were sent by the same source therefore the recipient node is unable to distinguish between legitimate and spoofed fragments. A malicious node can exploit this vulnerability to block reassembly of targeted packets such as connection establishment packets. This may result in disrupting legitimate traffic as well as depleting resources available to the victim node.  

\textbf{\textit {Buffer reservation attack:}}
The buffer reservation attack is closely linked to the fragment duplication attack and may be caused as a consequence of a successful fragment duplication attack. The buffer reservation attack also targets the vulnerability in the fragmentation mechanism employed by 6LoWPAN networks. As identified by \cite{XR1}, it leverages the fact that the recipient of a fragmented packet is unable to determine if all fragments will be received correctly. Therefore, a recipient node reserves a buffer space based on the information provided in the 6LoWPAN header with any additional fragments discarded. Taking advantage of this setting, a malicious node can send its victim single FRAG1 to reserve arbitrary buffer space thereby consuming scarce memory of the resource-constrained node. 

\textbf{\textit {Sybil and clone ID attack:}}
Sybil and Clone ID attacks are similar in that the objective of the attacker is to use spoofed logical identities within a network without deploying physical devices. In particular, for Clone ID attack, an attacker is aiming to use a victim’s logical identity within the network whereas in Sybil attack, the attacker aims to assume multiple logical identities within a network without deploying physical nodes. These logical identities may not be currently present in the network. A number of existing efforts such as  \cite{XR5,XR9} have identified these attacks for IoT and historically for wireless sensor networks.  

\subsection{Application specific threats}
In addition to the routing specific threats mentioned above, IoT infrastructures are susceptible to other types of threats such as application specific threats. Although routing forms an essential component of the IoT system, the IoT devices are expected to run application software required by the function envisaged to be performed. We categorize these threats as application specific and present them below.

\textbf{\textit {Denial of Service attack:}}
Historically, Denial of Service (DoS) attacks are targeted at making the victim unavailable for legitimate service. This can be achieved via flooding the victim with extraordinarily large volume of requests or by exhausting the resources such as memory and computational power available to the victim. Within IoT, the threat of DoS attack is two-folds: the victim can be part of the network under threat that an attacker wishes to make unavailable or the victim can be used as a zombie (stepping stone) to launch a Distributed DoS (DDoS) on a target IoT network. The significance of these threats within IoT systems have been identified by \cite{XR11, XR12,XR13}.

\textbf{\textit {Malicious code injection:}}
As identified by \cite{XR12,XR14}, malicious code injection is another application specific threat to IoT systems. The attacker, in this case, attempts to inject malicious code to get privileged access to the victim. Consequently, the attacker can damage the normal operation by causing threat to the data or to the network using one of the routing specific attacks described in the previous section.  

\subsection {Traditional attacks}
In addition to the above mentioned attacks, IoT systems are vulnerable to the existing attacks targeted at computer systems such as message interception, fabrication, modification, subversion and phishing etc. As with the routing-specific attacks, these attacks can also form a part of a more complicated/sophisticated attack.   

\section{Existing IDS Surveys for IoT}
\label{sec:existingsurveys}
An IDS is a hardware or software system which can be deployed at a system level or at the edge networks. It mainly serves to analyze the Internet packets and events (inbound and outbound) to identify malicious activities, and take actions according to the security policies of the network. IDS can be categorized into the following main approaches: misuse/signature based detection systems, behavioural or anomaly based detection, specification based intrusion detection and hybrid intrusion detection systems \cite{Butun,Meng,DEBAR1999805}.
The history of intrusion detection within IoT networks has its foundations within the Wireless Sensor Networks (WSN) where the focus has been on identifying and mitigating against threats misusing routing protocols. The routing protocols within such networks were optimized to work within a resource constrained environment and therefore prioritizing performance over security \cite{S3}. With the introduction of LoWPAN and RPL networks, sensor networks are now connected to the contemporary IP network resulting in expansion of the attack surface of such networks. Therefore, such networks are not only vulnerable to malicious attempts targeting routing protocols but also to the contemporary Internet-based attacks such as code injection, DoS and phishing - we present a bespoke attack model for IoT networks in the next section. We believe, the cutting edge efforts in IDS for IoT networks should take these considerations into account to mitigate against such malicious attempts.

The concept of IoT has evolved over the last two decades and therefore shares similarities with concepts such as Mobile Adhoc Networks (MANETs) and Wireless Sensor Networks (WSN). Consequently, a number of studies have been performed to review state of the art within intrusion detection for these paradigms. For instance, \cite{S5,S6,S7}, presented state of the art with respect to IDS for the MANETs, whereas \cite{S8,S9} have reviewed existing IDSs for WSNs. In this paper, we focus solely on the intrusion detection approaches for the IoT system in general and therefore applicable for its applications such as IoV.  

With respect to the review of the state of the art for the intrusion detection for the IoT  \cite{S1} and \cite{S2} represent the notable efforts. Gendreau et al. \cite{S1} define a term called \textit{Interaction Ability} of an IDS to assess the level of holistic detection intelligence. This parameter is defined as the ability of an IDS to interact with different service layers within the system i.e. Network Interface, Internet, Transport and Application layers. Therefore, the interaction ability of an IDS can have a maximum value of four. Authors have attempted to review seven recent intrusion detection systems against the interaction ability metric to identify respective efficiency. Although interaction ability is a useful indicator however we believe the intrusion detection landscape within IoT requires a rigorous analysis of existing efforts. For instance, one of the unique features of IoT systems is their limit to the resources available. Focusing on interaction ability alone ignores this significant characteristic and therefore although Jun and Chi \cite{R4} scores high (three) for interaction ability, indicating its efficiency to protect a IoT system. However, a deeper analysis of the system highlights that it is remarkably CPU intensive which will affect its suitability for IoT system. Therefore, we undertake rigorous analysis of the intrusion detection systems taking into account number of different parameters as explained in Section \ref{sec:reviews}.  

Furthermore, \cite{S2} present a recent survey of IDS research efforts for IoT with objective to identify leading trends, open issues, and future research possibilities. The authors classified the IDSs proposed in the literature according to the following attributes: detection method, IDS placement strategy, security threat and validation strategy. The authors also present a summary of security threats for a typical IoT system. Although the authors present a comprehensive system to analyze existing intrusion detection efforts, however, similar to \cite{S1}, this effort is agnostic of the impact of the intrusion detection operation with respect to its performance. This is significant in the case of IoT systems due to the limited resources for these devices such as CPU, memory, storage, bandwidth and battery. In view of this, we conduct a review which takes into account these attributes to present an exhaustive evaluation of existing approaches for intrusion detection in IoT systems.

In addition to the above, Zitta et al, present an effort for intrusion detection system for RFIDs in \cite{LR15} using raspberry Pi. As part of their study, the authors have conducted a comparative analysis of the existing intrusion detection and prevention systems (IDPS) using attributes such as location of the system, scalability and IPS function. This comparison is limited as i) it does not present an exhaustive selection of intrusion detection approaches for IoT devices, and ii) the comparison performed does not take into account the performance overhead of the intrusion detection approaches. 

Furthermore, intrusion detection within WSN is a well-established research domain which can be adopted within the IoT. However, most of these approaches are built on the assumptions that (i) there is no central management point and controller, (ii) there exists no message security, and (iii) nodes cannot be identified globally. The IoT has a novel architecture where the 6BR is assumed to be always accessible, end-to-end message security is a requirement \cite{5}, and sensor nodes are globally identified by an IP address. Besides these opportunistic features, an IDS for the IoT is still challenging since the things (i) are globally accessible, (ii) are resource constrained, (iii) are connected through lossy links, and (iv) use recent IoT protocols such as CoAP \cite{6}, Zigbee, RPL \cite{RPLRFC}, or 6LoWPAN \cite{2}. This analysis indicates the rationale for investigating and addressing the challenges for effective intrusion detection IoT.

\section{Review of Existing IDS Approaches for IoT}
\label{sec:reviews}
  
Through our research, we have identified that intrusion detection research for IoT systems can be categorized into: Anomaly, Signature, Specification, Hybrid, and Game based models. Therefore, we present review of existing efforts using these categorization.

\begin{figure*}
\begin{center}
 \includegraphics[width=170mm,height=145mm]{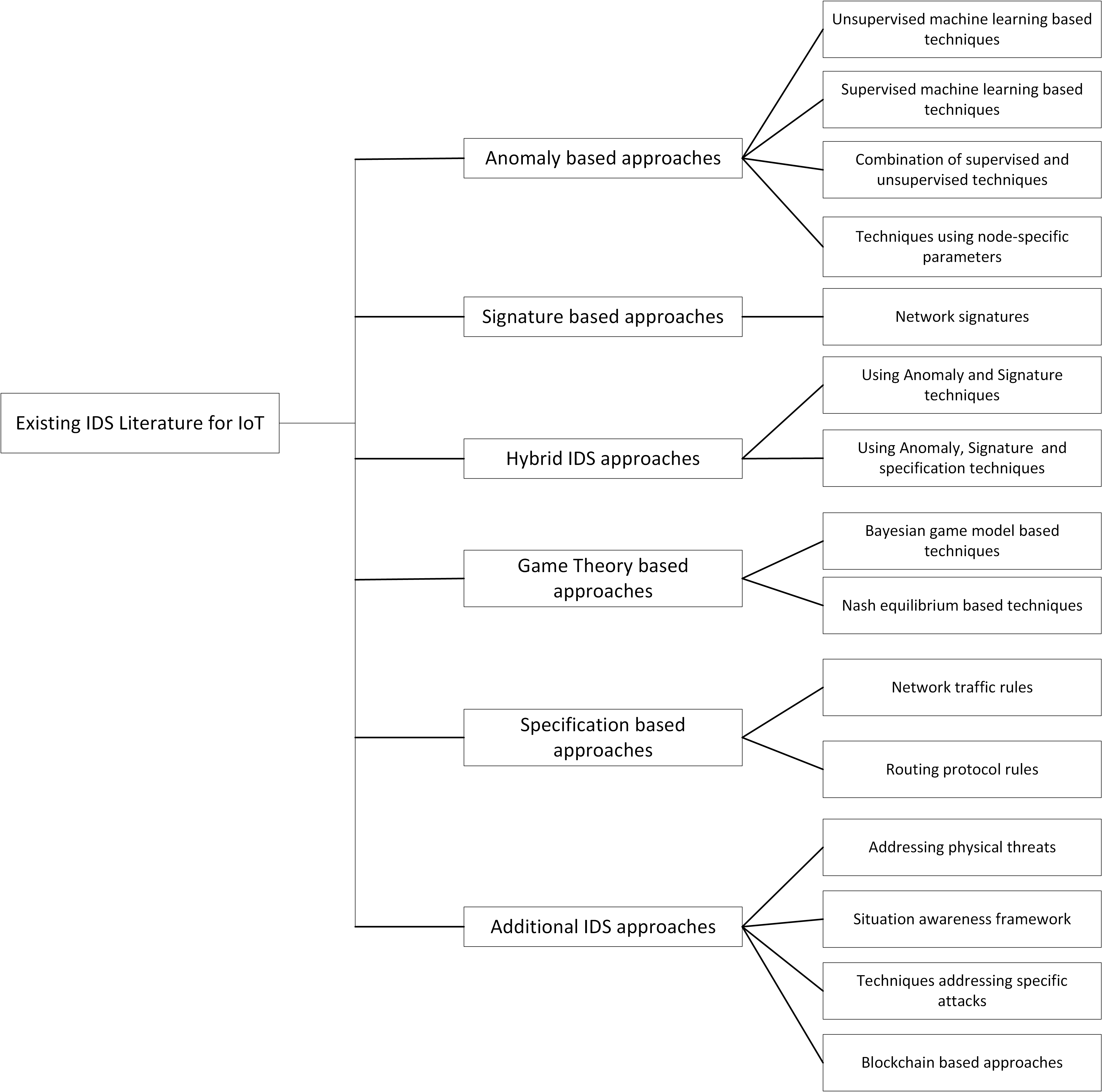}
  \setlength{\belowcaptionskip}{-10pt}
  \caption{Categorization of IDS for IoT systems}
  \label{fig:fig3}
  \end{center}
  \end{figure*}

\subsection{Anomaly based approaches to intrusion detection}

Nobakht et al. \cite{LR5} propose a host based IDS using Software Defined Technology (SDN) for smart homes. The authors have defined three basic requirements for an efficient IDS for IoT i.e. unobtrusive approach, negligible overheads, and scalability. In view of these requirements, the proposed approach uses sensors to host the intrusion detection module which is envisioned to monitor network traffic visible at the sensor device. The detection can be performed using a choice of detection modules i.e. signature, anomaly or specification based techniques. The authors claim that hosting the intrusion detection module within a sensor device alleviates the communication overhead. However, this choice increases the processing overhead at the sensor device simultaneously which is critical for such low powered devices.
 
In \cite{LR6}, Chordia and Gupta proposed an anomaly based IDS aiming to reduce false alarm rates and increase the detection efficiency using data mining techniques. The proposed system aims to monitor network traffic and uses techniques such as K-NN ,K-Means and Decision Table Majority Rule Based scheme. The authors have focused on four attacks namely  U2R ,R2R ,DoS and Probe. In the first phase, k-means clustering is used as pre-classification. In the next stage significant clusters are classified into attack classes by using K-NN. As a final step, Decision Table majority rule is used to categorize network events as malicious or non-malicious. The authors have used KDD99 dataset to evaluate the effectiveness of the approach which highlights the unavailability of data from an IoT system to aid more rigorous and proportionate evaluation.
 
Khan and Herrmann \cite{LR9} presented an effort where an IDS is designed and evaluated for IoT by using trust management mechanism that collects information about neighboring devices and their reputation. The authors investigate the patterns of normal use for the RPL protocol and use these patterns as a foundation to devise trust among the sensor nodes and the edge routers. Therefore, the proposed approach can be categorized as an anomaly based intrusion detection and is aimed at routing-specific attacks such as sinkhole, selective forwarding and version number. The trust management algorithms devised as part of this approach aim to develop trust and reputation values which are used by a border router to assess if a node is malicious or non-malicious.

An anomaly based distributed IDS for IoT is proposed in \cite{LR11} whereby each node monitors the working of nearby nodes for any abnormal activity. If some abnormality is detected then its packets are blocked and problem is reported to the parent node or root node by Distress Propagation Object (DPO). In the proposed system all information is managed locally within the nodes and parent nodes are notified if an anomaly occurs. It follows a Cross Layer approach so that unwanted packets are not processed at LowPAN Lower. The system has three subsystems i.e. Monitoring and Grading Subsystem (MGSS), Reporting Subsystem (RSS) and Isolated Subsystem (ISS). The MGSS serves as the primary component responsible for information collection, analysis and grading of a monitored threat whereas the RSS component is responsible for status update using DPO. The threats monitored by the MGSS are also reported to the edge router which can serve as a central alert gathering system with options to deploy correlation to further investigate or verify a threat.
 
Zhang et al. \cite{LR14} proposed a hierarchical and distributed IDS (SGDIDS) for smart grids which is applied to three layer communication network i.e. Home Area Network (HAN), Neighbourhood Area Network (NAN) and Wide Area Network (WAN). The WSN used in smart grids are prone to a variety of attacks like DoS or Man in the Middle (MIM). Power measurements can also be altered by different attacks. So Smart Grid without any IDS cannot produce expected operation. As the attack surface for a typical smart grid comprises of threats targeting the physical power system as well as the IT infrastructure supporting it, the SGDIDS is envisioned to address the cyber-physical nature of smart grids protecting against both these types of attacks. The intrusion detection architecture proposed as part of this effort has an IDS module developed at each of the three layers which monitor communications at each layer. The proposed system leverages classification algorithms such as support vector machine (SVM) and artificial immune system (AIS) in order to determine if an attack is occurring, what type of attack it is, and where it comes from in the communication system. The SGDIDS shares some similarities with a typical IoT network in that an IoT network can be divided into two levels i.e. IoT devices and the edge routers with the possibility to adopt SGDIDS approach for a hierarchical IDS. Furthermore, the attack surface for smart grids overlaps signifcantly with that for a typical IoT network therefore the results of this research are relevant within a generic IDS for IoT. 

\cite{LR16} presented a network anomaly based model for intrusion detection called Two Layer Dimension Reduction and Two Tier Classification TDTC model used in IoT backbone networks. The authors have focused on two specific attacks types i.e. User to Root (U2R) and Root to Local (R2L). The proposed system has advantages of multilayer classifier so has high detection rate with high performance, low false positive rates due to introduction of a refinement feature and decreased computational complications. The proposed model consists of a dimension reduction and a classification module. Dimension Reduction module is proposed to address the problems due to dimensionality which can increase the complexity of decision making and also leading to incorrect decisions. The authors have used both supervised and unsupervised reduction techniques to solve this issue. In particular, Linear Discriminant Analysis LDA(supervised) and Principal Component Analysis PCA (unsupervised) are used where PCA is used for feature selection and extraction, and LDA  provides fast and efficient IDS. In the proposed system Naive Bayes Classifier is used for anomaly detection which is normalized using certainty factor version of KNN. The authors have performed experimentation to evaluate the performance of TDTC identifying that it uses only two features of new mapped data instead of 4 and it is more efficient in detecting U2R, R2L, Probe and DoS attacks. TDTC is also reported to have reduced the false alarm rate from 6-3 \% to 5.56\%.

In \cite{LR18}, Summerville et al. presented a deep packet anomaly detection system involving feature selection conducted by pattern matching. It can be applied to either stateless or stateful configuration using sliding window operation which reduces the complexity. Two Internet enabled devices are considered which are: a weather station and an interactive networked video camera. They serve as sensor and actuator respectively. By experiments it was identified that IoT sensor detected 99.9\% of abnormal packets. IoT actuator detected a bit less as 92.9\%. Results showed that 64 by 32 detector has lower detection than higher size detectors. Also if we move from 2 dimensional feature to 3 dimensional feature then the accuracy improved from 47.7\% to 100\% per window and from 98.4\% to 100\% per packet. The proposed system can be implemented in hardware or software. It can be deployed into device interface or can be built into network appliances or firewalls. It results in low occurrence of false positives. Device specific traffic can be differentiated from other traffic through proper system.

A user centric approach is proposed in \cite{LR19} for security consisting of two main blocks i.e. a habit based approach for anomaly detection system and semantic based firewall for access control and security during communication. An algorithm is developed user’s habit modelling so that any change in the behaviour can be detected. The authors consider use of IoT devices within a private setting such as home networks with devices such as a wrist bracelet acting as a data storage device for an individual, a connected light bulb which indicates the on/off situation and a smart TV giving track of the programs/channels being watched. Within this context, the authors focus on the personal data collected and communicated by the devices. The authors use anomaly based approach for intrusion detection due to the unavailability of a signature database for attacks on IoT systems. The authors do not discuss details of the detection accuracy, performance efficiency, visibility for the intrusion detection and its placement. 

Yang et al. \cite{LR22} focuses on Node issue of IoT is addressed which can result in false data injection (FDI) attack which affects data aggregation. For this purpose Bayesian Spatial Temporal (HBST) model is described and an anomaly detection method is proposed to detect the compromised node at an early stage. In the proposed system divided difference filtering (DDF) based state estimation technique is used to detect false aggregated data. The system achieves high detection rate and low false positives rates with low detection samples. The proposed scheme works in static cluster base network nodes and follows an assumption that adversary is competent to compromise a subset of set of networks. It is capable of detecting compromised aggregator as long as the adversary launch the FDI attack with a probability of more than 58.7\%. Here a game theoretic model is presented overcome the problem of detection of false data injection. Simulation is done for effectiveness and efficiency using TOSSIM to measure the computation and storage overhead. According to the data model, a threshold decision mechanism is designed for data integrity. Performance of the proposed scheme is robust against interruption of attack behaviour and harsh environment.

\subsection{Signature based approaches to intrusion detection}
Kasinathan et al. \cite{LR3} presented an IDS framework for 6LoWPAN which was able to detect denial of service attacks by monitoring physical parameters of the device. The proposed IDS is included into an IoT network and monitors network traffic for both signatures and abnormal behaviour to identify malicious users. The intrusion detection component of the proposed system is implemented using an open source IDS Suricata which has complete IPv6 support, multithreading, automatic protocol detection and a built-in intrusion prevention system. The proposed system results in detecting flooding attacks with an increased detection rate where larger number of Mp nodes are in the same attack. Furthermore, Kasinathan et al. claim that the proposed IDS overcomes resource constraint problem which is one of the critical requirements for the sensor devices.
 
Forzin et al. \cite{LR4} proposed leveraging the Snort, contemporary signature based network level IDS, to establish a portable, easy to use and versatile intrusion detection  for IoT networks. The resultant IDS is packages within a Raspberry Pi so it can be transported with the device to any network the hardware travels to. By doing this, the device is not dependent on a centralized IDS which is a clear benefit of a host based intrusion detection system. In addition to this, the authors propose using these packaged RPi to work in collaboration to achieve detection of sophisticated attacks such as those targeting network topology. Under this setup, an IDS node can request data captured by neighboring nodes to perform rigorous analysis and reduce false positives. In view of the increasing threats for IoT systems with respect to volume and sophistication, this observation will be crucial to an effective intrusion detection for IoT systems.

Indre and Lemnaru \cite{LR10} proposed a modular architecture for intrusion detection is proposed which uses network traffic captured as part of the Data Capture Module. The captured network traffic is consumed by three modules employing different detection techniques i.e. signature, anomaly and specification based systems. The aim is to filter faulty packets by comparing packet with malicious header segment using static rules using signature based detection, anomaly based detection aims to identify abnormal data patterns and specification based detection focuses on botnet detection. The proposed approach is an example of approaches which use multiple type of detection techniques to achieve intrusion detection for IoT systems. However, the proposed approach envisages intrusion detection at the edge router level which compromises visibility of the system. Furthermore, although the proposed approach makes use of three different detection techniques however the mechanism to aggregate the outcome of these techniques is unclear. 

\subsection{Game based approaches to intrusion detection}
Wang et al. \cite{LR20} proposed an attack-defence game model is proposed to detect malicious nodes using repeated game approach. The authors claim to have developed a tree model which is used to formulate an optimal solution to the error detection problems. The authors focus on the detection strategy and the performance overhead caused by such a module on the individual sensor devices. This is achieved by developing a repeated game method where attackers and defenders can alter strategies to achieve maximum pay offs. The authors aim to achieve high detection accuracy without incurring significant performance overhead with experimental results supporting these claims. The authors do not discuss challenges such as placement of the IDS module, type of the detection engine and the level of visibility to it which can contribute towards the overall effectiveness of an IDS for M2M or IoT networks.

La et al. \cite{LR27} propose a honeypot based approach to improve defence against malicious attempts within an IoT infrastructure. In order to strengthen the defence, authors propose to use a game theoretic model with the defender using honeypot to deceive the attacker.
The proposed system focuses on scenarios where the attacker is not known to the defender which consequently means that the system does not take into account attempts such as Rank Attack which can be initiated by an insider node known to the system. The proposed system is designed and implemented in isolation to the specific challenges and requirements of the IoT infrastructures such as the resource constraints, interactions between different IoT devices, and the positioning of the system etc. The authors present the experimentation performed in Matlab which models the overall concept of the system however a Matlab based implementation can serve as a simulation of the proposed model agnostic of the IoT system. We believe this to be a limitation, as an effective intrusion detection mechanism for IoT systems should ideally take into account the unique characteristics of such systems as those explained above.

\onecolumn
{\footnotesize
\begin{longtable}[t]{|p{1cm}|p{1.2cm}|p{1.2cm}|p{1.6cm}|p{1.6cm}|p{2.5cm}|p{2.7cm}|p{2.5cm}|p{1.3cm}|}

\caption{Comparative analysis of existing IDS approaches for IoT}
\label{IDSanalysis}\\

\toprule
\textbf{Paper} & \textbf{Visibility} & \textbf{Time} & \textbf{Detection Engine} & \textbf{Architecture} & \textbf{Performance Overhead}& \textbf{Attack Types}                                   & \textbf{Detection Performance}      & \textbf{Scalability}\\

\midrule
\endfirsthead
\multicolumn{7}{l}{\footnotesize\itshape\tablename~\thetable:
continued from previous page} \\
\toprule
\textbf{Paper} & \textbf{Visibility} & \textbf{Time} & \textbf{Detection Engine} & \textbf{Architecture} & \textbf{Performance Overhead}& \textbf{Attack Types}                                   & \textbf{Detection Performance}      & \textbf{Scalability}\\
\midrule
\endhead

\midrule
\multicolumn{7}{r}{\footnotesize\itshape\tablename~\thetable:
Continued on next page} \\
\endfoot

\endlastfoot
\hline
Midi et al. \cite{LR1}  & Network  & Offline  & Hybrid    & Collaborative    & CPU usage:0.19\%, RAM usage (KB)=13978.62\%   & ICMP Flooding, SMURF                 & Detection Rate: 91\%, Accuracy: 100\%                   & Yes                                      \\[2ex] \hline
Fu et al. \cite{LR2}     & Network       & Realtime           & Specification based          & Distributed           & Latency: 4.12us, power consumption: 7.5w                  & Buffer overflow, probing, finger printing          & N/A            & Yes                         \\[2ex] \hline
Kasi.et al. \cite{LR3}      & Network          & Offline          & Signature based      & Distributed     & N/A                                                       & DoS             & Accuracy:100\%            & Limited         \\[2ex] \hline
Forzin et al. \cite{LR4}     & Network          & Combined         & Signature based       & Distributed       & CPU usage: 100\% for 70Mb/s, RAM usage: 475MB             & IP Spoofing     & N/A          & Limited                                  \\[2ex] \hline
Nobakth et al. \cite{LR5}     & Host          & Realtime        & Hybrid    & Distributed        & N/A     & Masquareding    & Accuracy: 94.25\%, Recall: 85.05\%      & Limited     \\[2ex] \hline
Chordia et al. \cite{LR6}     & Network      & Offline          & Anomaly based         & Centralised        & CPU usage: 49\%                                           & U2R, R2L, DoS, Probe     & Accuracy: 95.55\%, Detection rate: 93.67\%, FPR: 0.019  & No               \\[2ex] \hline
Jun et al.\cite{LR7}        & Network       & Realtime     & Specification based          & Centralised       & CPU usage: 48\%, RAM usage: 684MB, processing time: 368ms & Generic        & N/A             & No       \\[2ex] \hline
Khan et al. \cite{LR9}     & Network    & Offline      & Anomaly based        & Distributed            & N/A           & routing-specific attacks: sinkhole, selective forwarding and version number & For 300 nodes: FNR=26\%, FPR=47\%, Detection Rate= 50\% & Yes             \\[2ex] \hline
Indre et al. \cite{LR10}     & Network             & Offline       & Hybrid         & Centralised     & N/A               & Probing and DoS                              & Detection Accuracy=98.4\%       & No         \\[2ex] \hline
Thani. et al. \cite{LR11}        & Network     & Offline          & Anomaly based    & Distributed   & N/A     & Neighbouring node discrepancy                        & N/A                   & Yes                     \\[2ex] \hline
Amaral et al. \cite{LR12}    & Network         & Offline                & Specification based           & Distributed            & N/A                                  & Signature mismatching                     & N/A    & Yes                   \\[2ex] \hline
Zhang et al. \cite{LR14}      & Network     & Offline      & Anomaly based      & Distributed          & N/A                                                       & DoS, U2R, R2L      & FPR: 0.67, 0.7 and 1.3. FNR: 2.15, 21.02, 26.32\%       & Yes                 \\[2ex] \hline
Haddad et al.\cite{LR16}      & Network        & Offline      &  Anomaly based           &  Centralised       &     N/A             &      U2R and R2L                &   Detection accuracy: 81.97\% and FPR: 5.44\%           &  No                  \\[2ex] \hline
Summ. et al. \cite{LR18}          & Network    & Offline        &        Anomaly based      &  Distributed    & N/A	    & Worm propagation, tunneling, SQL code injection       & Detection accuracy: 100\%    & Yes                           \\[2ex] \hline
Tamani et al. \cite{LR19}       & Network    & Offline    &  Anomaly based     & Distributed     & N/A      & Privacy Threats     &  N/A    &  Yes                                        \\[2ex] \hline
Wang et al. \cite{LR20}    & Network      & Offline         & Game theory   & Distributed     & Energy consumption: avg 2.0J for 300 nodes, Energy consumption: avg 2.0J for 300 nodes   & N/A    & Detection accuracy: avg. 80\%           &Yes                                          \\[2ex] \hline
Yang et al. \cite{LR22}       & Network      & Offline         & Anomaly based & N/A       &  Energy consumption: 8,48mJ        &                                       &  N/A           &        N/A                      \\[2ex] \hline
Sedje. et al. \cite{LR23} & Network   & Offline     &      Hybrid             &   Distributed                   &     Efficiency: 2s, energy consumption: approx 3000 mj for 300 nodes               &       DoS        &    Detection accuracy 92\% for large number of nodes. FPR: 3\%                                                     &   Yes                                       \\[2ex] \hline
Raza et al. \cite{LR28}    & Network   & Realtime           &   Hybrid           &      Distributed          &    efficiecny: >150000mj for 64 nodes      & sinkhole and selective forwarding attacks       &    TPR: approx 80\% for 30 nodes on avg.           &   Yes             \\[2ex] \hline
Le et al. \cite{LR29}     & Network      & Realtime              &    Specification based              &     Hybrid               &      energy consumption: 202J, power consumption: 6.3\% increase (1.2mW)              & Rank, sinkhole and neighbour attacks             & TPR: 100\%, FPR: apprix: 3\% avg        & Yes           \\[2ex] \hline
Mayza. et al. \cite{LR32} & Network    & Offline           &     Other       &      Distributed               & N/A               &  version number attacks                 &  FPR: 0\% for some simulations              &  Yes \\[2ex] \hline 
Arshad et al. \cite{LR37} & Both    & Offline           &     Hybrid       &      Collaborative               & energy consumption: avg 8.5mW for 1, 10, 100 and 1000 packets/sec               &  Generic                 &  N/A              &  Yes \\[2ex] \hline  
\hline
\end{longtable}
}
\twocolumn

\subsection{Specification based approaches to intrusion detection}
Fu et al. \cite{LR2} propose an IDS for Internet of Vehicles (IoV) i.e. an open and integrated network system connecting human intelligence, vehicles, things, environments and the internet. Authors explain fundamental requirements for an IDS for IoV i.e host protection via detection, constrained resources, and real-time detection. Through our research we identify these to be shared by a typical IoT system.  Due to these requirements, conventional IDS is not feasible as it incurs significant overhead with respect to required resources and processing time. In order to fulfil these requirements, an FPGA based IDS is proposed which can work in real time and applicable to IoV security however it limits versatility of the approach as it is unclear if the system can be translated for other hardware. An extended NFA called Link NFA is proposed to eliminate the transition problem of Multistride NFA. The rulesets are adopted from the existing IDS like Snort, Bro and L7 Filter. In \cite{LR7} a Complex Event Processing  (CEP) based approach to intrusion detection for IoT is proposed with the aim to provide low latency and real time processing of security events. Although the use of CEP for intrusion detection is not novel, it is a rather innovative concept within intrusion detection for IoT and this work can be considered as a means to assess the feasibility of CEP within IoT. The proposed system is developed as a rule based IDS which collects network traffic data from the edge router, extracts events from this data and performs intrusion detection. Although the discussion of the approach is not detailed, we believe CEP can be explored further to address sophisticated or coordinated attacks within IoT which are a combination of simple attack steps.

Amaral et al. \cite{LR12} present a network based intrusion detection system for internet-connected WSN which share close similarities with a typical IoT network. The authors adopt a distributed approach to their design in that the IDS system is deployed as a module on randomly selected devices called watchdogs (sensor nodes as well as edge routers) over a WSN. Although the IDS module is deployed on individual devices, it does not monitor system status and only monitors network traffic visible at the node. However, as the watchdogs devices are heterogenous in location and function, this may have an impact on the types of threats encountered at a specific device. Therefore, each watchdog device is pre-configured with custom monitoring rules which are designed to detect specific threats expected to be encountered by a device with all the watchdog devices reporting to a centralised Event Management System (EMS). We believe that the distributed approach adopted by the authors is symmetric with the inherent characteristics of a typical IoT network however the role of EMS is limited to information collection. This constrained role of EMS limits the IDS’s ability to monitor the overall state of a network and therefore is unable to detect coordinated and sophisticated attacks.

Le et al. \cite{LR29} presented one of the early efforts to establish an IDS for IoT where authors proposed a host-based IDS for LoWPANs using Contiki OS and 6LowPAN \cite{3,4}. The IDS is able to perform detection based on the information at the node level and then transmit data to some centralized system for further analysis. The detection system performs detection using information collected from individual nodes and does not consider the information from other nodes in the network. The system does not show effective detection under the distributed denial of service attack that not only overwhelms the device but also congests the communication channel between nodes and the centralized system.

\subsection{Hybrid approaches to intrusion detection}
Midi et al. \cite{LR1} proposed a self-adapting, knowledge-driven expert Intrusion Detection System (KALIS) is introduced which can change its performance after evaluating its efficiency. It is focused on network features and its protocols to improve detection efficiency. KALIS automatically collects the information about the features and configure most suitable detection technique. All the KALIS components runs independently and it supports a wide variety of mediums and related protocols. The existing IDS for IoT deploy an independent IDS on each device comprising the IoT network but the disadvantage is that each IDS is related to only one device and does not have security details of other devices. In KALIS the attack scenarios in IoT are considered and their differences from other domains than IoT and relationship is studied between different networks and device features and attacks. Therefore, KALIS is concerned with the feature details of a network and its components consequently deciding upon most effective IDS for that scenario. The attacks considered by KALIS are variants of DoS attacks i.e ICMP Flood and SMURF which produce similar symptoms i.e. high amount of ICMP Echo Reply Messages directed to affect the machine.

Sedjelmaci et al. \cite{LR23} a game theoretic technology is proposed to activate anomaly detection technique to have a balance between detection and false positives and achieve energy consumption. The authors propose a novel lightweight anomaly based intrusion detection technique for IoT taking into account the resource constraints of the IoT devices. The overall system in fact combines both signature and anomaly based detection with the anomaly based engine only activated when a previously unknown attack signature is encountered. The authors adopt a game theoretic approach based on Nash Equilibrium (NE) to determine the equilibrium state in which the IDS agent will activate its anomaly detection technique to train, classify and build a rule related to a new attack’s signature. Furthermore, with respect to placement of IDS, both the signature and anomaly based detection systems are based on individual sensor devices. The authors explain this decision to be motivated by the objective to reduce the communication overhead based on the assumption that the communication overhead is more resource hungry as compared to computational overhead. However, this is expected to have significant computational overhead in the case of zero day attacks. Furthermore, the proposed system is limited in its visibility to the events occurring on the individual nodes and therefore limits its ability to effectively detect of sophisticated attacks which may be composed of multiple steps.

Raza et al. \cite{LR28} proposed Svelte is proposed which is a novel IDS for IoT. The main focus is on Routing attacks like Spoofing, Sinkhole and Selective Forwarding. It is Lightweight and comply with the processing capabilities of constrained nodes. The authors have designed their system to be hybrid i.e. combining both signature and anomaly based detection, and placing detection function at both the sensors devices and the gateway router. Svelte has three main modules i.e. 6LoWPAN Mapper (6 mapper), Intrusion Detection component and Distributed Mini Firewall.The centralized modules have two corresponding lightweight modules in each constrained node. The first module provides mapping information to the 6BR so it can perform intrusion detection.The second module works with the centralized firewall. Each constrained node also has a third module to handle end-to-end packet loss. Although Svelte presents a hybrid approach, attempting to adopt a best of both approach however there are considerations that should be taken into account. For instance, although anomaly based intrusion detection systems have the advantage of better detection accuracy for zero-day attacks, they are typically resource hungry which can be a bottleneck when implementing them on resource constrained sensor devices. Furthermore, known attacks such as Rank attack, Sinkhole and spoofing are well established leading to an established signature for their detection. However, attacks such as multi-stage or zero-day require analysis from a wider perspective, analyzing behavioral and usage patterns which makes a case for using anomaly based IDS. Therefore, an IDS for IoT should take these factors into consideration when making choices such as the type of IDS, placement and the visibility of the data.

Sheikan et al. \cite{LR37} presents a recent effort to address intrusion detection with IoT systems particularly focusing at the constrained resources available at the sensor devices. The authors present a collaborative approach where IDS modules are implemented at both device and edge-router level to improve visibility, detection rate and to reduce false positives. the authors propose signature based IDS at the node level due to its performance efficiency and anomaly based detection at the edge-router to enhance the detection accuracy. The approach is novel in that it proposes a innovative solution to achieve high detection accuracy whilst taking into account the limited resources available at the sensor devices. 
 
\subsection{Additional intrusion detection efforts within IoT}
Daramas et al. \cite{LR8} an enhanced and safe home based IDS HIVE is proposed having 3 parts i.e. a sensor manager, firebase as cloud database and user authentication service, and android application for monitoring, configuring and remote notification. The proposed system is especially design for a smart home with focus on detecting physical intrusion into a home. HIVE aims to detect a physical intrusion by using three sensors i.e. an infrared sensor to detect motions, a magnetic switch sensor to detect status of a door or window and a load cell sensor to detect pressure such as footsteps.
 
The resource constraints in general and the limited available of power in particular are one of the challenges for any IDS within IoT. To this end, Gendreau \cite{LR25} seek to address this challenge by using an enhanced measurement of situation awareness in the IoT. The authors highlight the importance of awareness of state of monitoring system and propose a framework to enhance the energy efficiency of a self-reliant management and monitoring WSN cluster head selection algorithm. Although the experimental results demonstrate positive results for the approach however it is limited to assessing the energy efficiency for a cluster head selection algorithm and therefore require further efforts to assess its impact on the monitoring capabilities of individual nodes.
 
Mayzaud et al. \cite{LR32} authors proposed a distributed system architecture for detecting the version number attacks in RPL-based networks and identifies malicious nodes. Furthermore, a number of intrusion detection system architectures have been developed in \cite{LR34,LR35} for the resource-constrained 6LoWPAN devices based systems focusing on the sinkhole and selective-forwarding attacks (well-known attacks within 6LoWPAN networks).
 
Golomb et al. \cite{LR36} present an innovative approach, CIOTA, to intrusion detection for IoT leveraging blockchain technology. The proposed approach is comprised of local agents and a central component which coordinates information (alerts) received from these agents. Authors use blockchain technology to achieve assurances about the authenticity of alerts generated by local agents.

\section{Privacy Implications of Intrusion Detection Systems}
\label{sec:privacyimplications}
Intrusion detection systems are widely used to protect the devices and end-users from the malicious actors. The performance of these systems often depends on the type of data and architectural setup they use while preventing the attacker from misusing the private information of victims for the frauds. The usage of data for the intrusion detection introduces the challenges of security and privacy. Therefore, we discuss the privacy implications of intrusion detection systems in this section.

\subsection{Type of data}
Intrusion detection systems can be categorized on the basis of network data they used for detecting the malicious actors. Identification of correct data type not only affects the performance of the system locally and globally, but also has an impact on the privacy of the users. For instance, if the detection system uses IP-address to analyze the behavior of device, then it can easily traceback to the owner of this IP-address. The challenge in this regard is two fold: 1) identifying the data type that provides optimum performance in terms of detection rate, and 2) ensure protection of the privacy of network users. The existing intrusion detection systems use two types of data for detecting the malicious actors i.e. 1) application layer data logs, and 2) network traces data. The first type is data originated at the application level and is normally associated with specific type of data set. This type of data can provide information about the device architecture and can help in fingerprinting the malicious and non-malicious devices as normally malicious devices do not follow the specific set of standards. The second data type is the IP traces of the traffic and contains much more details about the behavior of the devices. Another data type that can be used is the content or payload of the information exchanged among devices such as a temperature reading, a web-page, or meta data. 
\subsection{System architecture}
An intrusion detection system for IoT can operate in two modes i.e. 1) as the standalone system, or 2) as a collaborative system. The stand-alone detection systems rely on the traffic patterns observed locally within the network domain or Internet service provider. These systems work independently within a service provider network. The stand-alone systems do not have any information about the behavior of its users in other domains thus can easily be circumvented by stealth techniques and smart attacker such as by simply controlling the attack traffic to one domain but target large number of domains simultaneously. An effective intrusion detection system should consider the collective behaviour of nodes generating traffic across different domains thus building the collaborative network. \\
The collaborative solutions can be grouped in two types: 1) centralized - where alert information from the domain collaborators is reported to the the centralized system which classifies the behavior of traffic sender by analyzing traffic patterns from multiple domains, or 2) The distributed or decentralized settings - where alert information from each service provider is shared and processed in a completely distributed fashion without a centralized coordinator.

The major challenge towards the design of collaborative intrusion detection system is regarding privacy protections for the data used for detection. The domain provider or Internet service provider are reluctant to share operational data of their users with each other as it risks privacy of their customers. A centralized trusted aggregation can overcome the problem of privacy if the centralized repository assures cooperating domain that their provided information would not be misused and disclosed to any one. Furthermore, use of cryptographic methods or addition of noise data can be considered to anonymize user data, however this is expected to substantially increase the network load and computation time.
\subsection{Analysis}
In this section, we analyze the privacy implications of IoT intrusion detection system for two important features, the data type used for collaboration, and the system architecture.

Firstly, the system architecture of the detection system determines how data is transferred by the entities in the detection systems. Standalone detection system which are installed on the user device (e.g. IoT device) or installed at the edge router (entry point router to smart home or the company network) operate locally by recording data from the single source, only use data from the single source, therefore does not have high performance accuracy. The collaborative system, however, operates in two modes the centralized \cite{LR6} \cite{LR16} \cite{LR10} architecture and the distributed architecture \cite{LR18} \cite{LR19} \cite{LR20}. In the centralized setup, the centralized trusted setup is envisaged to protect the privacy and integrity of the data provided to it. However, it may not be ideal as the attacker has to compromise only one device to breach the privacy of all collaborators. Furthermore, the centralized system introduces the challenge of single point failure, which may inhibit efficient collaboration in case of failure. 

The transfer of data to other parties or centralized system has a risk of privacy. In this setting, the collaborating device can operate in four settings: 1) transfer all the raw data to the centralized system or other devices that then process all the data for meaningful decision. This setting does not have any privacy assurance as data is exposed at other entity whilst also increasing computational workload, 2) transfer processed data for instance exchanging traffic statistics of host or IP-address , however it still carries privacy threat but without requiring significant additional resources, and  3) the encrypted exchange of data. This setting assures privacy-preservation but requires extensive computation and communication overhead for the exchange of encrypted data.

Overall, privacy is an important feature which should be afforded by intrusion detection system especially within a collaborative system. However, our analysis of the existing literature reveals that the research community has not given much attention to privacy preserving collaboration among the IoT domains or IoT devices. This may be due to resource-constrained nature of IoT devices which limits alert information and cryptographic processing of data due to computational overheads.

\section{Analysis of Current IDS Approaches}
\label{sec:analysis}
In order to conduct a rigorous and methodical analysis of contemporary literature presented in the section \ref{sec:reviews}, we have applied a thorough criteria with metrics which are significant for effective intrusion detection for IoT. The individual element of the criteria along with a brief explanation are presented below. The comparative analysis of existing approaches for these criteria is presented in Table 1.
\begin{itemize}

\item \textbf {Placement:} As with the contemporary computing systems, the placement of an intrusion detection system is crucial as it determines the level of visibility it can offer to the activities within the monitored system. For instance, a network based IDS is limited to monitoring the network traffic originated or destined for the monitored host and therefore cannot monitor any process subversion or privilege escalation within the monitored host. 
\item \textbf {Detection Time frame:} One of the important characteristic of IoT systems is the dynamic nature of the system with the participating nodes following an adhoc pattern. Therefore the time-frame of detection becomes even more important with the objective to detect an attack as soon as possible to avoid spreading infection to wider devices. 
 
\item \textbf{Detection Engine:} An IDS can utilize a variety of detection engine such as anomaly, signature and game based etc. The choice of detection engine has two-fold impact i.e. i) it can affect the ability of an IDS to detect attacks and ii) it impacts the performance overhead incurred by the engine. For instance, although signature based IDS have been identified to be resource efficient, they do not have the ability to detect zero-day attacks.
 
\item \textbf{Architecture:} The system architecture of the intrusion detection system specifies how the detection system carried out its detection functions. The system architecture not only affects the detection accuracy and performance but also affects user privacy. The standalone detection system mainly operates at the local machine or the device thereby susceptible to extended detection time because of the non-availability of enough data and stealthy nature of the attacker. A collaborative architecture utilizes  data from different sources e.g. IoT devices or network devices within the same or different organization. It can improve the detection accuracy however it introduces the challenge of privacy of the data shared between the entities. Furthermore, with regards to detection accuracy and performance, a typical IoT system is comprised of a number of sensor devices arranged into a local network such as a LoWPAN and an edge router which manages communication between local network and the Internet. Within this context, existing approaches can be categorized based on the location of the IDS module with distributed referring to IDS module implemented on local sensor nodes and centralized referring to IDS module implemented at the edge router.  

\item \textbf{Performance Overhead:} A typical IoT device is constrained with respect to resources available for compute processes such as intrusion detection. Therefore, we believe performance overhead caused by an IDS is one of the important criterion and can be measured in the form of energy consumption or CPU usage by the IDS. 

\item \textbf{Attack types:} We have presented a comprehensive discussion about the potential attacks within an IoT system supported by an attack model. These different types of attacks can be detected at different levels (network or host) and using different approaches such as anomaly, signature based etc.

\item \textbf{Detection Performance:} Detection performance represents the rate of with which an IDS can successfully detect a malicious attempt. It is one of the fundamental attributes of an IDS as it can be directly aligned to its effectiveness.

\item \textbf{Scalability:} Typically, the number of devices within an IoT system is significantly higher compared with contemporary systems. In order to address the significant number of devices involved, the scalability of the IDS is an important criteria.  
\end{itemize}

\section{Open Challenges and Future Directions}
\label{sec:challenges}

In this section, we highlight some future research directions which require further investigation and efforts to improve overall security of an IoT system. 

\begin{enumerate}
\item \textbf{Constrained resources:}
A typical IoT device has limited resources such as small processing power, low storage capacity, and limited battery power. Within this context, non-resource efficient intrusion detection system would drain the resources of the IoT system and its devices. Therefore, it is important to have a Intrusion detection system that fulfills two important characteristics: 1) any IDS should not incur significant computational and communication overhead, and 2) IDS should achieve high detection accuracy. 
In particular, the use of anomaly based detection systems \cite{LR22} \cite{LR19} \cite{LR18} requires more resources than the signature based detection systems while having a tradeoff between detection accuracy and overheads. For instance, anomaly detection is particularly effective against previously unknown attacks, but is expected to incur significant performance overhead. As an attempt to explore opportunities within this context, we have formulated a collaborative intrusion detection system in \cite{LR37} which aims to use both anomaly and signature based detection engines to achieve performance efficiency without compromising detection accuracy.  \\
Our analysis has revealed that many of the existing systems have not been designed for resource-constrained devices, however these approaches mainly focused on increasing the detection accuracy with small false positive. We believe there should be trade-off among three important factors 1) high detection accuracy, 2) small overheads, and 3) privacy-preservation. Furthermore, dedicated efforts are required to devise approaches considering resource constraints that primarily focus on the energy consumption agnostic of resources and memory consumption.

\item \textbf{Multi-stage attacks:}
A typical attacker normally carries an attack in multiple stages.  Such sophisticated attacks are termed as \textit{Multi-stage attacks}, and are common attack mechanisms for traditional and emerging computing systems such as IoT. Existing detection systems for IoT solely focus on the detection of individual threats agnostic of potential relationships between them. We believe the dynamic nature of the IoT systems makes the challenge of multi-stage attack detection non-trivial requiring explicit efforts to address it. Jun and Chi \cite{R4} represent one of the initial effort which recognize and explicitly seek to detect relationships between different malicious incidents. However, it represents a limited effort and further work is required to address detection and protection against multi-stage attacks within the IoT systems.

\item \textbf{Device protection:}
As identified by \cite{DCMS}, one of core issues with respect to the security of IoT systems is the security of the device as \textit{"it is often neglected by the manufacturers and usually an afterthought"}. The lack of protection at device level within such systems has resulted in significant security attacks such as Mirai botnet in 2016 \cite{Mirai2016} and its more recent versions such as Brickerbot \cite{bricker} and Reaper \cite{Reaper2018}. In order to protect against such threats device-level security measures are paramount which can protect IoT systems. One such measure can be an effective intrusion detection installed within the IoT device. Through our research we have identified efforts such as \cite{LR5} and \cite{LR37} which propose to develop intrusion detection capability within the device however these efforts are generally limited in that these require further refinement to take into account unique characteristics of IoT devices such as those explained earlier in this section. 

\item \textbf{Large-scale attacks:}
With the widespread adoption of IoT systems, the number of IoT devices are increasing exponentially with some estimates predicting more than 50 billion IoT devices by the year 2020. The impact of this growth on securing the IoT system is is two-folds; firstly, the enormous scale makes IoT systems a lucrative target for the malicious actors, and secondly, it also presents IoT systems as a resource which can be used to launch a large-scale attacks. An example of such attacks is the botnets i.e. Mirai botnet and Brickerbot launched a Distributed Denial of Service (DDoS) which compromised the Domain Name System (DNS) service \cite{Mirai2016}. Moreover, due to the nature of the IoT systems, routing attacks are typically contagious i.e. affecting all the devices within a LoWPAN.  
These attacks demand a holistic approach to the intrusion detection which is able to monitor and detect the state of the overall network as well as the individual devices. 

\item \textbf{Limited experimentation and evaluation:}
In order to assess the effectiveness of intrusion detection efforts, rigorous experimentation is required for integrating multiple dimensions of the evaluation. Although, experiments have been conducted to demonstrate the effectiveness with respect to detection accuracy and false positive rate, but this evaluation is performed without using appropriate simulation software or hardware to replicate a real-life IoT setting. For instance, a number of efforts have used KDD 99 dataset \cite{KDD99} within an isolated environment to conduct experimentation, however it has limitations: 1) the KDD 99 dataset does not accurately reflect the current threat types prevalent for the IoT systems and, 2) conducting evaluation in an isolated environment prohibits taking into account important factors such as resource constraints of a typical IoT device. These issues require explicit efforts to improve state of the art with respect to the evaluation of intrusion detection schemes within IoT systems.Further, we believe that research into dedicated honeypots for IoT systems is required and will be significant in facilitating thorough evaluation of future intrusion detection approaches. 

\item \textbf{Unavailability of accurate data:}
Through our research, we have identified unavailability of real data from an IoT system as one of the bottlenecks to achieve rigorous evaluation. In absence of such data, a number of current approaches have used KDD99 datasets which contains network traffic data for contemporary computing systems. We believe further research into dedicated honeypots for IoT systems is required and will be significant in facilitating thorough evaluation of future intrusion detection approaches. 

\end{enumerate}

\section{Conclusions}
\label{sec:conclusions}
The emergence of IoT has been led by the extraordinary evolution of the sensor devices and communication technologies such as Zigbee, WSN and 6LoWPAN. 
Consequently the volume and variety of security threats for such systems have increased manifold highlighting the importance of an efficient intrusion detection system. This paper has presented a comprehensive review of existing efforts within this domain aiming to identify open challenges and future directions. The paper has provided new focus on the performance overhead, energy consumption and privacy implications incurred by existing approaches. Highlighting challenges surrounding these aspects, this review has attempted to enthuse researchers to address key challenges identified in this article to achieve effective intrusion detection for IoT systems.

\bibliographystyle{IEEEtran}
\bibliography{refs}

\begin{thebibliography}{10}
\bibitem{bricker}
I.~Thomson, ``"forget mirai – brickerbot malware will kill your crap iot
  devices".''
\end{thebibliography}
\end{document}